\documentclass[fleqn,usenatbib]{mnras}

\usepackage{newtxtext,newtxmath}

\usepackage[T1]{fontenc}

\DeclareRobustCommand{\VAN}[3]{#2}
\let\VANthebibliography\thebibliography
\def\thebibliography{\DeclareRobustCommand{\VAN}[3]{##3}\VANthebibliography}

\usepackage{graphicx}	

\usepackage{amsmath}	
\usepackage{amssymb}	

\usepackage[usenames,dvipsnames]{xcolor}

\title[Variability in transiting white dwarfs]{Long-term variability in debris transiting white dwarfs}

\author[Aungwerojwit et al.]{
Amornrat Aungwerojwit$^{1}$\thanks{E-mail: amornrata@nu.ac.th},
Boris T. G\"ansicke$^{2,3}$, 
Vikram, S. Dhillon$^{4,5}$,
Andrew Drake$^{6}$, \newauthor
Keith Inight$^{2}$,
Thomas~G. Kaye$^{7}$,
T.~R. Marsh$^{2}$,
Ed Mullen$^{8}$,
Ingrid Pelisoli$^{2}$,
Andrew Swan$^{2}$\\
$^{1}$Department of Physics, Faculty of Science, Naresuan University, Phitsanulok, 65000, Thailand\\
$^{2}$Department of Physics, University of Warwick, Coventry, CV4 7AL, UK\\
$^{3}$Centre for Exoplanets and Habitability, University of Warwick, Coventry, CV4 7AL, UK\\
$^{4}$Department of Physics and Astronomy, University of Sheffield, Sheffield S3 7RH, UK\\
$^{5}$Instituto de Astrof{\'i}sica de Canarias, E-38205 La Laguna, Tenerife, Spain\\
$^{6}$Department of Astronomy, California Institute of Technology, 1200 E. California Boulevard, Pasadena, CA, 91125, USA\\
$^{7}$Foundation for Scientific Advancement, Sierra Vista, AZ, USA\\
$^{8}$Sycamore Canyon Observatory, Vail, AZ, USA
}
\date{Accepted XXX. Received YYY; in original form ZZZ}

\pubyear{2021}

\begin{document}
\label{firstpage}
\pagerange{\pageref{firstpage}--\pageref{lastpage}}
\maketitle

\begin{abstract}
Combining archival photometric observations from multiple large-area surveys spanning the past 17 years, we detect long-term variability in the light curves of ZTF\,J032833.52$-$121945.27 (ZTF\,J0328$-$1219), ZTF\,J092311.41+423634.16 (ZTF\,J0923+4236) and WD\,1145+017, all known to exhibit transits from planetary debris. ZTF\,J0328$-$1219 showed an overall fading in brightness from 2011 through to 2015, with a maximum dimming of $\simeq0.3$\,mag, and still remains $\simeq0.1$\,mag fainter compared to 2006.  We complement the analysis of the long-term behaviour of these systems with high-speed photometry. In the case of ZTF\,J0923+4236 and WD\,1145+017, the time-series photometry exhibits vast variations in the level of transit activity, both in terms of numbers of transits, as well as their shapes and depths, and these variations correlate with the overall brightness of the systems.  Inspecting the current known sample of white dwarfs with transiting debris, we estimate that similar photometric signatures may be detectable in one in a few hundred of all white dwarfs. Accounting for the highly aligned geometry required to detect transits, our estimates imply that a substantial fraction of all white dwarfs exhibiting photospheric metal pollution from accreted debris host close-in planetesimals that are currently undergoing disintegration.
\end{abstract}

\begin{keywords}
minor planets -- asteroids:  general -- planets  and  satellites:  physical  evolution -- planetary systems -- white dwarfs
\end{keywords}



\section{Introduction} 
Practically all known planet hosts will evolve eventually into white dwarfs, and large parts of the various components of their planetary systems~--~planets, moons, asteroids and comets will survive that metamorphosis \citep{sackmannetal93-1, mustill+villaver12-1, adamsetal13-1, verasetal14-3, payneetal16-1, caiazzo+heyl17-1, mustilletal18-1, martinetal20-1, maldonadoetal21-1}.  Rich observational evidence for such evolved planetary systems at white dwarfs exists in the form of photospheric contamination from the accretion of planetary debris \citep{zuckermanetal03-1, koesteretal14-1}, infrared excess and metallic emission lines from compact dusty and gaseous debris discs \citep{zuckerman+becklin87-1, gaensickeetal06-3}, transits from both debris \citep{vanderburgetal15-1, vanderboschetal20-1} and planets \citep{vanderburgetal20-1}, and spectroscopic detections of planetesimals \citep{manseretal19-1}, and giant planets \citep{gaensickeetal19-1, schreiberetal19-1}. 

Among all those features, white dwarfs accreting the debris of tidally disrupted rocky bodies \citep{debesetal02-1, jura03-1} are most common-place. Yet, our understanding of the full sequence of the processes involved is still limited, beginning with the gravitational scattering of these bodies onto orbits which take them into the tidal disruption radius of the white dwarf \citep{bonsor+wyatt12-1, frewen+hansen14-1, stephanetal17-1, smallwoodetal18-1, antoniadou+veras19-1}, the actual disruption and subsequent circularization of the debris into a compact circumstellar disc \citep{debesetal12-2, verasetal14-1, verasetal15-1, brownetal17-1, malamud+perets20-1} and the further evolution of these discs \citep{metzgeretal12-1, kenyonetal17-1, kenyonetal17-2, miranda+rafikov18-1}. Although observational estimates of the disc life times are $\sim10^4-10^6$\,yr \citep{girvenetal12-1, cunninghametal21-1}, there is growing evidence of variability in the optical \citep{gaensickeetal18-1, wilsonetal14-1, manseretal16-1, dennihyetal18-1, gentile-fusilloetal21-1}, and infrared emission from these discs \citep{xuetal14-2, xuetal18-2, swanetal19-1, swanetal20-1}, that these discs are not static systems, but dynamically active. 

The first genuine signature of an ongoing disruption event was the discovery of multiple transits from debris at WD\,1145+017 with orbital periods of $\simeq4.5$\,h \citep{vanderburgetal15-1}, consistent with quasi-circular orbits \citep{gurrietal17-1, verasetal17-2} near the tidal disruption radius for a rocky body. The system also exhibits strong photospheric metal contamination as well as an infrared excess and gaseous absorption lines from the circumstellar debris \citep{vanderburgetal15-1, xuetal16-1, fortin-archambaultetal20-1}~--~as such, WD\,1145+017 was the textbook example that confirmed the general framework initially laid out by \citet{jura03-1}. Intense monitoring of WD\,1145+017 since its discovery shows that the amount of occulting debris varies dramatically on time scales of months to years, demonstrating the ongoing dynamic activity within the system \citep{gaensickeetal16-1, garyetal17-1, rappaportetal18-1}.

In stark contrast, the debris transits in ZTF\,J013906.17+524536.89 (hereafter ZTF\,J0139+5245), the second system discovered, recur every $\simeq107$\,d \citep{vanderboschetal20-1}, highlighting that the tidal disruption processes are likely to be more complex \citep{verasetal20-1, malamud+perets20-2}, or that indeed thermal destruction could play a key role \citep{shestakova+serebryanskiy23-1}. 

\citet{guidryetal20-1} announced the identification of five additional systems that were good candidates for displaying debris transits. One of these, ZTF\,J032833.52$-$121945.27 (hereafter ZTF\,J0328$-$1219) has been followed up in more detail by \citet{vanderboschetal21-1}, who identified two orbital periods, 9.94\,h and 11.2\,h. Finally, \citet{farihietal22-1} discovered debris transits with an orbital period of 25\,h in WD\,1054$-$226. It is now clear that actively disintegrating planetesimals can be found spanning a wide range in orbital parameters, and levels of dynamical activity. 

\begin{figure}
\includegraphics[width=\columnwidth]{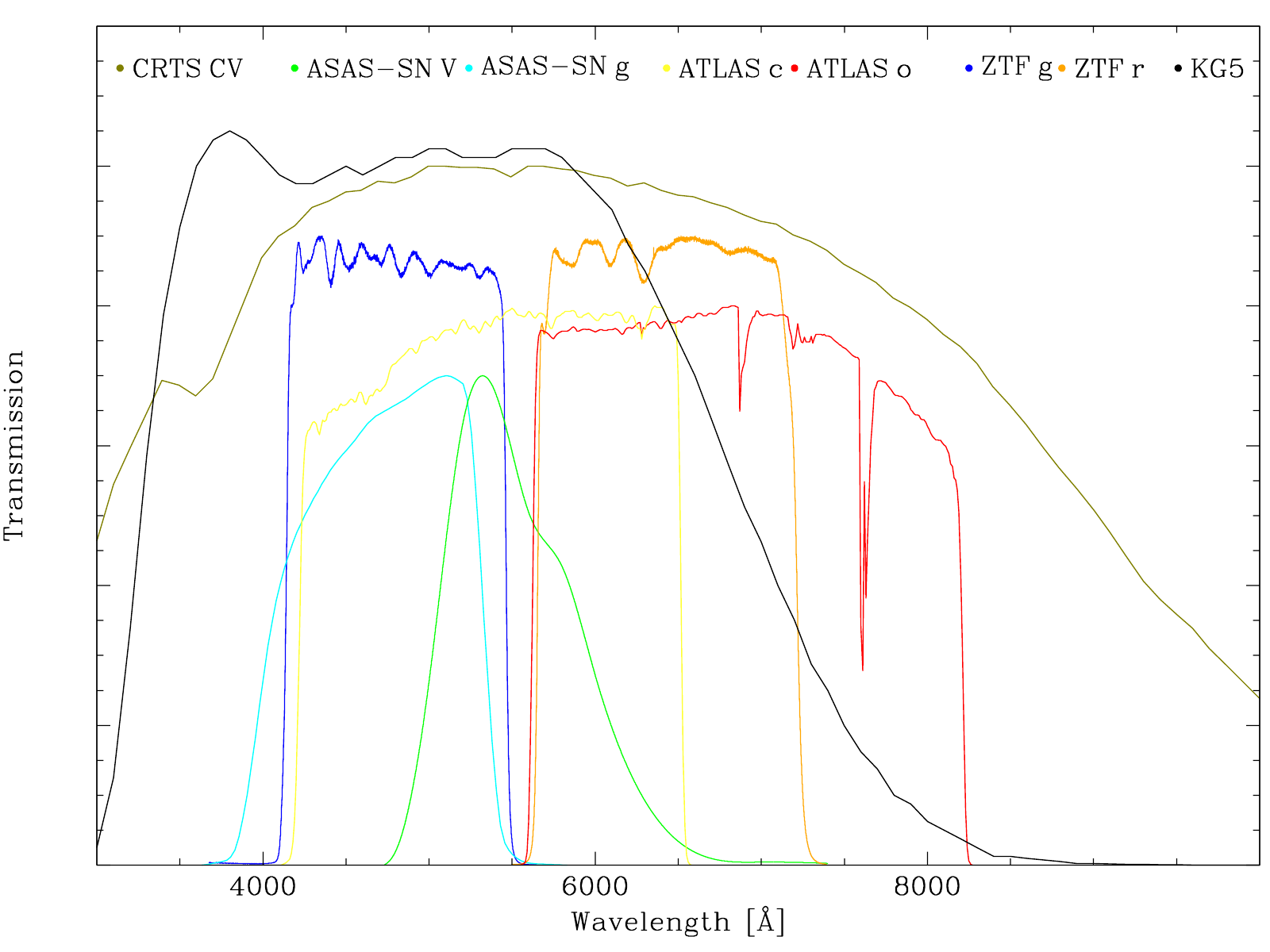}
\caption{The transmission curves of the photometric surveys used for this study. The transmission curves for CRTS, ZTF, and ATLAS were retrieved from the Spanish Virtual Observatory filter service \citep{rodrigo+solano20-1} apart from the KG5 filter, which was obtained from the HiPERCAM/ULTRACAM/ULTRASPEC filter database\protect\footnotemark. For the ASAS-SN filters, we show representative $V$ and $g$-band curves. The curves have been scaled vertically for clarity, i.e. are not reflective of the absolute transmission of the different surveys.}
\label{fig:bandpasses}
\end{figure}

\footnotetext{http://www.vikdhillon.staff.shef.ac.uk/ultracam/filters/filters.html}

Whereas ongoing searches for additional systems will further map out the parameter space of orbital periods, eccentricities, and multiplicity of the fragments causing the transits, it is equally important to better characterise the eight systems identified so far, in particular to explore the evolution of their debris. Here, we make use of the data from large-area photometric surveys and detailed time-series photometry to investigate the long-term variability and the changes in transit activity and morphology of three white dwarfs with debris transits: ZTF\,J0328$-$1219, ZTF\,J0923+4236, and WD\,1145+017.

\section{Observations}
\subsection{Time-domain surveys}
We retrieved the sparse, long-term photometry of ZTF\,J0328$-$1219, ZTF\,J0923+4236, and WD\,1145+017 from the following wide-field time-domain surveys (Table\,\ref{tab:survey_obslog}).

\begin{table}
\caption{\label{tab:survey_obslog} Summary of the time-domain survey photometry used for the analysis of the three white dwarfs exhibiting transits from circumstellar debris.}
\setlength{\tabcolsep}{0.95ex}
\begin{flushleft}
\begin{tabular}{llccr}
\hline\noalign{\smallskip}
Star & Survey & Date range &  Filter & Epochs \\
\hline\noalign{\smallskip}
\textbf{ZTF\,J0328$-$1219} & CRTS    & 2005-12-01 to 2015-12-02 & Clear & 293\\
                  & ASAS-SN & 2012-01-29 to 2018-11-29 & $V$   & 551\\
                  & ASAS-SN & 2017-09-16 to 2023-08-11 & $g$   & 2030\\
                  & ATLAS   & 2015-10-02 to 2023-08-14 & $o$   & 2565\\
                  & ATLAS   & 2015-08-09 to 2023-08-15 & $c$   & 709\\
                  & ZTF     & 2017-12-20 to 2023-02-16 & $g$   & 363\\
                  & ZTF     & 2017-11-16 to 2023-02-19 & $r$   & 375\\\hline
\textbf{ZTF\,J0923+4236}   & CRTS    & 2005-12-08 to 2016-04-18 & Clear & 486\\
                  & ASAS-SN & 2012-10-22 to 2018-04-12 & $V$   & 68\\
                  & ASAS-SN & 2017-10-29 to 2023-06-15 & $g$   & 598\\
                  & ATLAS   & 2015-10-05 to 2023-06-05 & $o$   & 2036\\
                  & ATLAS   & 2015-10-21 to 2023-06-09 & $c$   & 573\\
                  & ZTF     & 2017-12-20 to 2023-05-01 & $g$   & 617\\
                  & ZTF     & 2017-10-28 to 2023-05-01 & $r$   & 832\\\hline
\textbf{WD\,1145+017}      & CRTS    & 2005-12-02 to 2016-05-09 & Clear & 771\\
                  & ASAS-SN & 2012-02-16 to 2018-11-13 & $V$   & 285\\
                  & ASAS-SN & 2017-11-30 to 2023-07-16 & $g$   & 1403\\
                  & ATLAS   & 2016-01-21 to 2023-07-30 & $o$   & 1795\\
                  & ATLAS   & 2016-01-13 to 2023-06-16 & $c$   & 580\\
                  & ZTF     & 2017-12-19 to 2023-04-30 & $g$   & 227\\
                  & ZTF     & 2017-12-03 to 2023-04-29 & $r$   & 341\\
\hline
\end{tabular}
\end{flushleft}
\end{table}

\subsubsection{Catalina Real Time Transient Survey}
The Catalina Real Time Transient Survey (CRTS; \citealt{drakeetal09-1}) uses filter-less photometry obtained with three telescopes: the Catalina Sky Survey telescope (CSS, 0.7\,m aperture, covering $\mathrm{Dec}=-25$\,deg to $+70$\,deg, \citealt{larsonetal03-1}), the Mt. Lemmon Survey telescope (MLS, 1.5\,m aperture, covering $\mathrm{Dec}=-5$\,deg to $+5$\,deg) and the  Siding Springs Survey telescope (SSS,  0.5\,m aperture, covering $\mathrm{Dec}=-80$\,deg to $0$\,deg). The survey strategy was optimised for the detection of near-Earth objects, obtaining typically four 30\,s exposures per night, with limiting magnitudes of $V\simeq19.0$, 19.5, 21.5 for the SSS, CSS, and MLS observations. Object detection and aperture-based photometry were obtained using the SExtractor \citep{bertin+arnouts96-1} package. A public data base\footnote{http://nesssi.cacr.caltech.edu/DataRelease/} provides access to observations obtained between 2005 and 2014. We used a CRTS-internal data base that provides additional observations until May 2016.

\subsubsection{All-Sky Automated Survey for Supernovae}
The All-Sky Automated Survey for Supernovae (ASAS-SN; \citealt{shappeeetal14-1, kochaneketal17-1}) consists of a global network of 0.14\,m telescopes obtaining nightly all-sky photometry, using $V$- and $g$-band filters, with a limiting magnitude of $V\simeq17-18$\,mag. Each ASAS-SN epoch consists of three dithered 90\,s exposures. We retrieved the ASAS-SN light curves via positional queries using the aperture photometry service provided by the SkyPatrol\footnote{https://asas-sn.osu.edu/}. We only included detections in our analysis, i.e. excluded epochs which only provided an upper limit on the brightness of our targets.

\subsubsection{Asteroid Terrestrial-impact Last Alert System}
The Asteroid Terrestrial-impact Last Alert System (ATLAS; \citealt{tonryetal18-1, smithetal20-1}) uses a global network of 0.5\,m telescopes to provide nightly ($-50<\mathrm{Dec}<+50$) or bi-nightly (in the polar regions) photometry using broad-band cyan ($c$-band) and orange ($o$-band) filters. The data is obtained as four 30\,s exposures spaced out over about an hour to maximise the efficiency for detecting near-Earth asteroids. We obtained the ATLAS light curves by using the forced photometry service\footnote{https://fallingstar-data.com/forcedphot/}, including proper-motions from \textit{Gaia} Data Release~3 \citep{gaiaetal20-1}. We only included in our analysis detections with photometric uncertainties $<0.1$\,mag.

\subsubsection{Zwicky Transient Factory}
The Zwicky Transient Factory (ZTF; \citealt{bellmetal19-1, mascietal19-1}) uses the 48\,inch (1.2\,m) Palomar telescope taking 30\,s exposures primarily in the $g$- and $r$-bands, with less frequent observation in the $i$-band, with a typical limiting magnitude of $r\simeq20.5-21.0$\,mag. ZTF carries out multiple surveys in parallel: the public surveys re-visit the Northern sky every three days and the Galactic plane every day, and private surveys are used for other experiments, such as observing dedicated fields several times per night for transient searches.  We retrieved the $g$-band and $r$-band PSF photometry within ZTF Data Release~19 from the Infrared Science Archive (IRSA) service, using the recommended \textsc{BAD\_CATFLAGS\_MASK=32768} bitmask to filter out bad quality data.

\begin{table}
\caption{\label{tab:tnt_obslog} Log of the TNT photometric observations, Exp. is the exposure time, Mag. the magnitude in the KG5 filter, and Trans. the average transmission.}
\setlength{\tabcolsep}{0.95ex}
\begin{flushleft}
\begin{tabular}{lcccccc}
\hline\noalign{\smallskip}
Date & UT &  Filter & Exp. & Frames & Mag. & Trans.\\
     &    &         & (s)  &        &       \\
\hline\noalign{\smallskip}
\multicolumn{6}{l}{\textbf{ZTF\,J0328$-$1219}} \\
2020 Dec 23 & 12:09-18:11 & KG5 & 3.0 & 4621 & 17.08 & 0.91 \\
2020 Dec 24 & 11:53-18:10 & KG5 & 3.0 & 5536 & 17.03 & 0.92 \\
2021 Dec 01 & 13:08-19:57 & KG5 & 5.0 & 4030 & 16.99 & 0.92 \\
\noalign{\smallskip}
\multicolumn{6}{l}{\textbf{ZTF\,J0923+4236}} \\
2020 Dec 23 & 18:53-23:18 & KG5 & 7.0 & 1908  & 18.04 & 0.88 \\
2020 Dec 24 & 18:21-23:17 & KG5 & 7.0 & 2026  & 18.02 & 0.87 \\
2021 Mar 12 & 12:20-19:37 & KG5 & 10.0 & 2328 & 17.90 & 0.90 \\
2021 Mar 13 & 12:25-19:43 & KG5 & 10.0 & 2236 & 17.89 & 0.92 \\
2022 Mar 27 & 13:58-19:28 & KG5 & 7.0 & 2466  & 17.85 & 0.93 \\
2022 Mar 30 & 14:17-18:24 & KG5 & 7.0 & 1835  & 17.84 & 0.89 \\
\noalign{\smallskip}
\multicolumn{6}{l}{\textbf{WD\,1145+017}} \\
2016 Jan 18 & 17:14-23:13 & KG5 & 3.0 & 3325 & 17.57 & 0.81 \\
2016 Feb 10 & 17:38-23:06 & KG5 & 3.0 & 3089 & 17.52 & 0.80 \\
2017 Feb 08 & 18:17-23:10 & KG5 & 3.0 & 4815 & 17.53 & 0.85 \\
2017 Mar 02 & 17:52-22:58 & KG5 & 3.0 & 4893 & 17.56 & 0.87 \\
2018 Feb 24 & 14:52-19:53 & KG5 & 3.0 & 4920 & 17.51 & 0.88 \\
2019 Feb 04 & 18:38-23:16 & KG5 & 3.0 & 4546 & 17.41 & 0.94 \\
2020 Jan 09 & 17:54-23:20 & KG5 & 5.0 & 3443 & 17.43 & 0.89 \\
2021 Feb 22 & 18:40-23:01 & KG5 & 5.0 & 2682 & 17.49 & 0.93 \\
2022 Mar 26 & 14:36-21:05 & KG5 & 5.0 & 4072 & 17.45 & 0.95 \\
2023 Feb 20 & 16:28-21:50 & KG5 & 6.2 & 2787 & 17.47 & 0.94 \\
%
\hline
\end{tabular}
\end{flushleft}
\end{table}

\subsubsection{Survey cross-calibration}
\begin{figure*}
\centering{\includegraphics[width=1.0\textwidth]{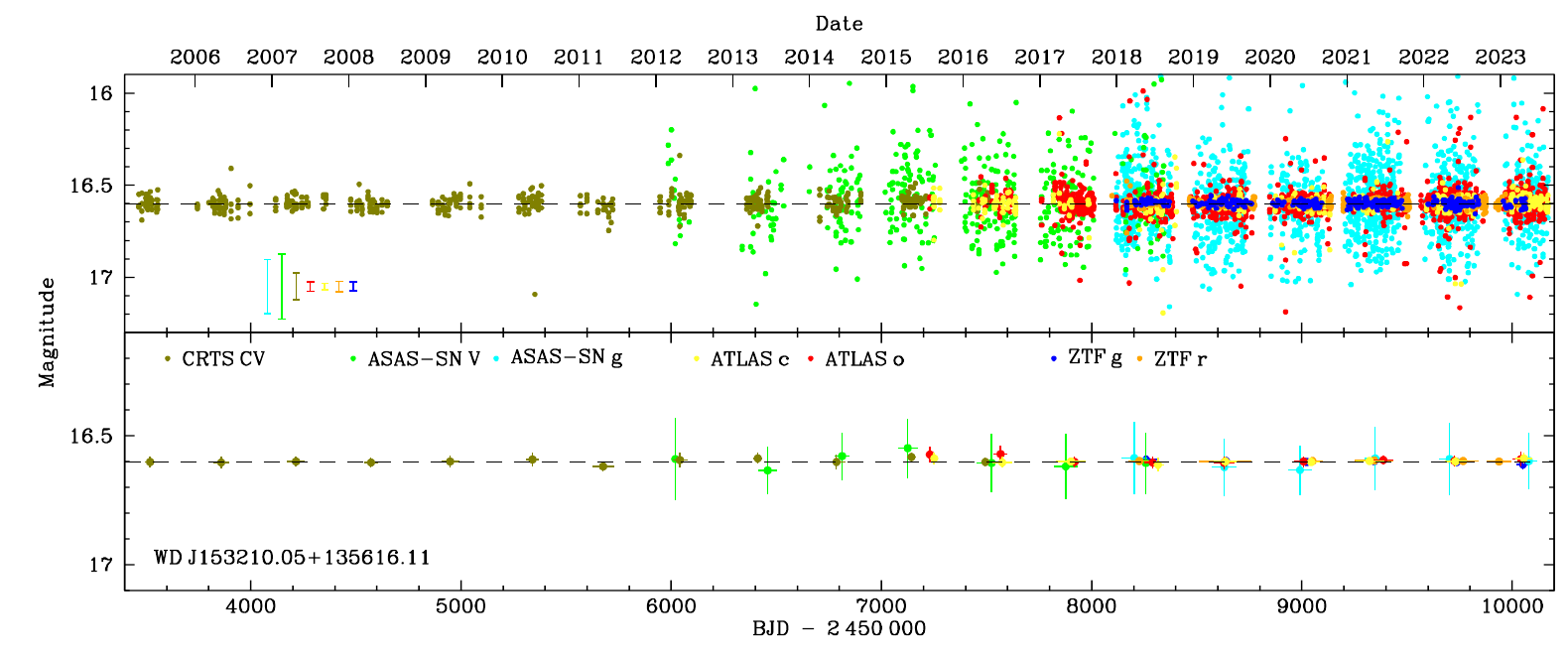}}
\caption{Long-term light curve of the white dwarf WD\,J153210.05+135616.11, which we used as reference star for ZTF\,J0328$-$1219. The individual observations are shown in the top panel. Median uncertainties of the individual measurements are displayed for each of the survey band passes in the lower left corner. The light curves of the different surveys have been offset such that their median magnitudes match that of the ZTF $g$-band data. The bottom panel shows the seasonal averages for each survey. The dashed lines in the bottom and top panels indicate the median brightness computed from the individual CRTS measurements obtained during the first two seasons (2006--2007). For this non-variable white dwarf, the seasonal averages agree within their respective uncertainties.}
\label{fig:wdj1532_long}
\end{figure*}

Our goal was to use all available survey data to establish the long-term light curves of the three debris-transiting systems, which requires to account for the offsets between the different filter band passes (Fig.\,\ref{fig:bandpasses}). For that purpose, we selected single white dwarfs from the catalogue of \citet{gentile-fusilloetal21-1} which have Pan-STARRS1 \citep{chambersetal16-1} $g-i$ colours as close as possible to those of the transiting systems, and retrieved their survey data as well. We then computed the median magnitudes of the reference white dwarfs across all six filter band passes, and scaled the CRTS, ASAS-SN, ATLAS, and ZTF $r$-band data to match the median magnitude of the ZTF $g$-band data. The resulting scaled long-term light curve for WD\,J153210.05+135616.11 (WD\,J1532+1356, also known as GD\,184) is shown in Fig.\,\ref{fig:wdj1532_long}. Checking the survey data of other white dwarfs with nearly identical $g-i$ colours, and using them as mutual reference stars, we established that they were non-variable. We found that this approach works robustly with the exception of the ASAS-SN $V$ and $g$-band data, which can show residual offsets with respect to the other surveys of $\simeq\pm0.1$\,mag. Despite various investigations, we were unable to unambiguously identify the source of these offsets. We note that the targets of our study are near the faint limit of ASAS-SN, and that data for an individual system is taken at multiple telescopes. Slight differences in the response functions of the different telescopes or the treatment of sky background and scattered light may contribute to the small offsets we found. We adopted WDJ\,145452.89+084640.44 as reference star for WD\,1145+017, and given the large-amplitude variability of ZTF\,J0923+4236, we decided to use scale the median magnitudes of that system itself to that of its ZTF $g$-band data.  

One caveat in the interpretation of variability detected in the joint-up survey photometry is that white dwarfs exhibiting transits caused by planetary debris might change both in brightness and colour, which would indeed be expected if a fraction of the circumstellar material is in the form of micro-metre sized dust. Simultaneous multi-wavelength studies of WD\,1145+017 concluded that the transit depths did not show noticeable changes across the optical to infrared wavelength ranges \citep{zhouetal16-1, crolletal17-1, xuetal18-1, xuetal19-2}. However, \citet{hallakounetal17-1} reported \textit{shallower} transits at shorter wavelengths ($u-r\simeq-0.05$\,mag) in WD\,1145+017. The authors suggested that this ``blueing'' is related to a reduction in the strength of absorption lines from circumstellar gas during the transits (see also \citealt{izquierdoetal18-1}). Given that most of the photometry discussed here does not cover wavelengths $<4000$\,\AA, where the majority of strong metal absorption lines are located, the effect of circumstellar gas should not strongly affect our analysis.

\subsection{\label{sec:ultraspec} ULTRASPEC fast photometry}
During 2016--2023, we obtained high-speed photometry of ZTF\,J0328$-$1219, ZTF\,J0923+4236, and WD\,1145+017 using the frame-transfer camera ULTRASPEC \citep{dhillonetal14-1} mounted on the 2.4\,m Thai National Telescope (TNT) on Doi Inthanon. We used a KG5 short-pass filter which cuts off red light beyond 7000\,\AA, and exposure times varying between 3.0  and 10.0\,s, depending on the brightness of the objects and the observing conditions, with less than 15\,ms dead time. The details of observations are provided in Table~\ref{tab:tnt_obslog}. All data-sets were reduced using HiPERCAM data reduction pipeline\footnote{https://github.com/HiPERCAM/hipercam}. The light curves of these ZTF\,J0328$-$1219, ZTF\,J0923+4236, and WD\,1145+017 are shown in Fig.\,\ref{fig:0328_tnt}, Fig.\,\ref{fig:0923_tnt}, and Fig.~\ref{fig:1145_tnt}, respectively.

We normalised each TNT light curve by computing the median flux of the 50 brightest data points, and setting that flux level to unity, and report the average transmission with respect to that level for each light curve in Table\,\ref{tab:tnt_obslog}. In cases where light curves exhibit clean out-of-transit stretches, this definition would allow a measurement of average absorption by the transits, however, as we will see later, in many cases the TNT light curves show continuously varying levels of flux, and hence our normalisation is only relative to the brightest moments captured by a given light curve.

\subsection{Liverpool Telescope}
We obtained sparse photometry of ZTF\,J0923+4236 with the Liverpool Telescope (LT) from 2020 December 19 to 2021 January 13, using the IO:O camera. We used the Bessel $B$-band filter which provides high throughput over the range $\simeq3800-4900$\,\AA\ and the default detector binning of $2\times2$. We collected a total of 88 60\,s images. The LT data are provided in a reduced (bias-corrected and flat-fielded) format, and we extracted the photometry of ZTF\,J0923+4236 using the pipeline of \citet{gaensickeetal04-1}. 

\subsection{Time-series monitoring of WD\,1145+017}
Intense time-series photometric observations of WD\,1145+017 have been obtained since the discovery of transits in 2015 using a range of observatories. We report here the average transmission measured from individual light curves obtained at Hereford Arizona Observatory (HAO) using a 16" telescope (see \citealt{rappaportetal16-1} for details), at Sycamore Canyon Observatory (SCO) using a 20" telescope, and at Raemor Vista Observatory (RVO) using a 1.1\,m fully automated Dahl-Kirkam with an Apogee camera equipped with a back-illuminated E2V chip. Exposure times were typically in the range $30-60$\,s and the SCO and RVO data were reduced in a standard fashion using the AstroimageJ software. 

\begin{figure*}
\centering{\includegraphics[width=1.0\textwidth]{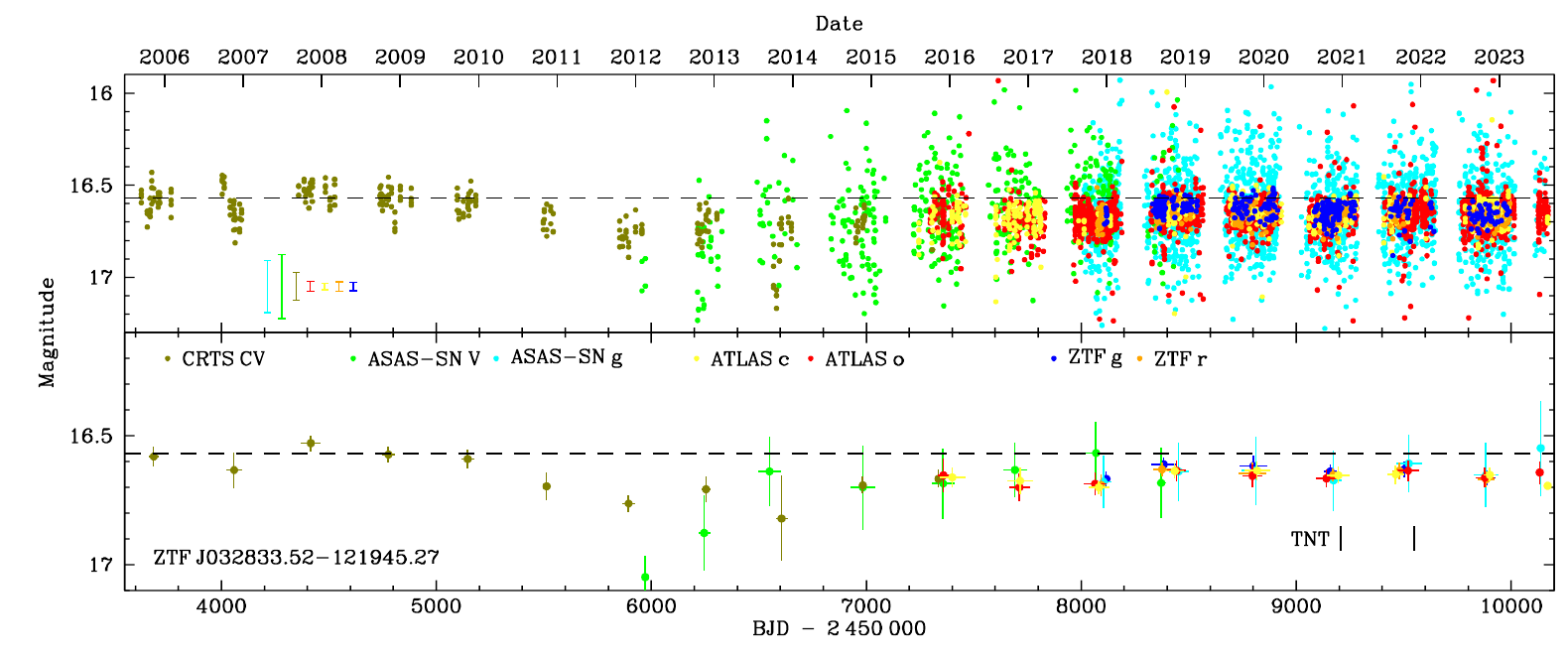}}
\caption{Long-term light curve of ZTF\,J0328$-$1219, all symbols have the same meaning as in Fig.\,\ref{fig:wdj1532_long}. We have used WD\,J153210.05+135616.11 as reference star to compute offsets between the different surveys. The CRTS data show that the system began to gradually dim around 2011, reaching a minimum brightness around 2013 to 2014, after which it recovered in brightness, though, not quite reaching the level of the early CRTS seasons. The times at which we obtained TNT/ULTRASPEC photometry are indicated by tick-marks in the bottom panel, the corresponding light curves are shown in Fig.\,\ref{fig:0328_tnt}.}
\label{fig:0328_long}
\end{figure*}

\begin{figure}
\centering{\includegraphics[width=0.5\textwidth]{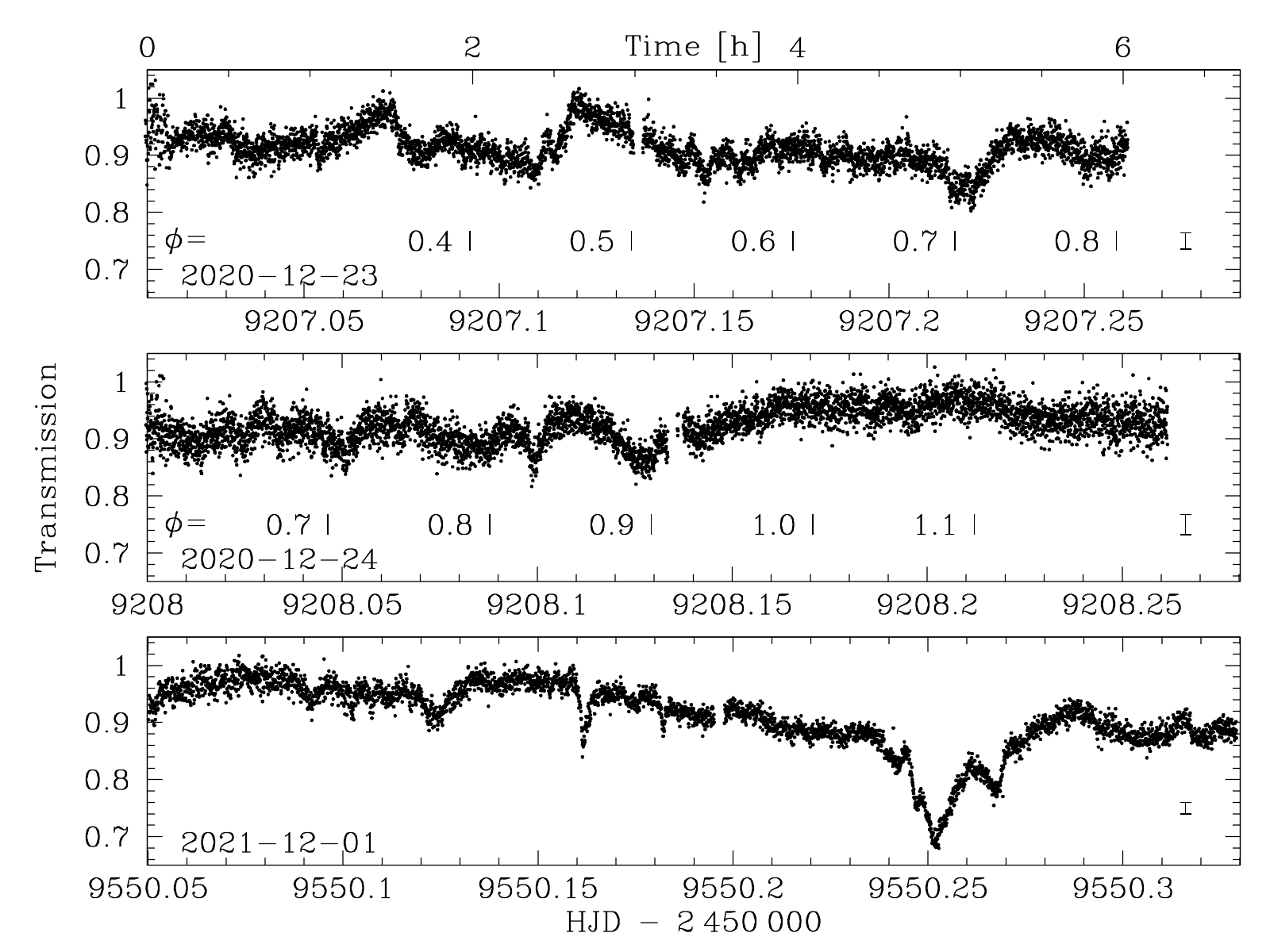}}
\caption{TNT light curves of ZTF\,0328$-$1219 displaying both sharp, narrow transits and broader, more complex structured absorption features. The median values of the photometric uncertainties are illustrated by the error bar in the bottom-right of each panel. Note that each observing run is shorter than the likely orbital periods detected by \citet{vanderboschetal21-1}, 9.94\,h and 11.2\,h. We computed orbital phases for the light curves shown in the top two panels, using the ephemeris of the 9.94\,h period given by \citet{vanderboschetal21-1}, see their fig.\,2 (their ephemeris is too uncertain to be applied to our 2021 December 1 data shown in the bottom panel). Orbital phases are labelled and indicated by tick marks below the light curves. The morphology of the light curve undergoes changes over a small number of cycles, possibly due to the beat with the secondary period of 11.2\,h or intrinsic variability. Nevertheless, some structures are reminiscent to those seen in the light curves of \citet{vanderboschetal21-1}, taken between 2020 December 12 and 15, e.g. the dip at $\phi\simeq0.7$ and the broad minimum near $\phi\simeq0.9$.}
\label{fig:0328_tnt}
\end{figure}

\section{Results}

\subsection{ZTF\,J032833.52$-$121945.27}
ZTF\,J0328$-$1219 was discovered as a white dwarf exhibiting transits of circumstellar debris by \citet{guidryetal20-1}, and the subsequent analysis of \textit{TESS} and ground-based photometry by \citet{vanderboschetal21-1} led to the identification of two periodic signals, 9.34\,h and 11.2\,h, which the authors interpreted as the orbital periods of debris fragments. 

The CRTS light curve of ZTF\,J0328$-$1219  (Fig.\,\ref{fig:0328_long}), which exhibits gaps due to the seasonal visibility of the star, displays a slight dimming beginning in $\simeq2011$. The seasonal average brightness continued to drop in the following years, ending up $\simeq0.3$\,mag fainter in 2014, compared to the first season in 2006, after which CRTS recorded the system brightening again. The reference star, WD\,J1532+1356, shows a constant brightness across the 17 years of survey coverage, with consistent seasonal averages across the different surveys (Fig.\,\ref{fig:wdj1532_long}). As an additional check, we investigated the possibility of the long-term brightness variation seen in the CRTS data of ZTF\,J0328$-$1219 being related to instrumental effects by extracting the CRTS light curves of three nearby ($<1.5$\,arcmin) stars, i.e. which were observed in the same field as ZTF\,J0328$-$1219. The light curves of these stars remain constant in brightness throughout the operations of CRTS, confirming that the dimming of  ZTF\,J0328$-$1219 is intrinsic to the system. This faint state is confirmed independently by the first seasons of the ASAS-SN $V$-band data, which then shows the brightness of ZTF\,J0328$-$1219 to recover. Throughout the years 2016 to 2023, ZTF, ASAS-SN and ZTF show the system at nearly constant magnitude, with consistent seasonal magnitudes across the different surveys. 

\begin{figure*}
\centering{\includegraphics[width=1.0\textwidth]{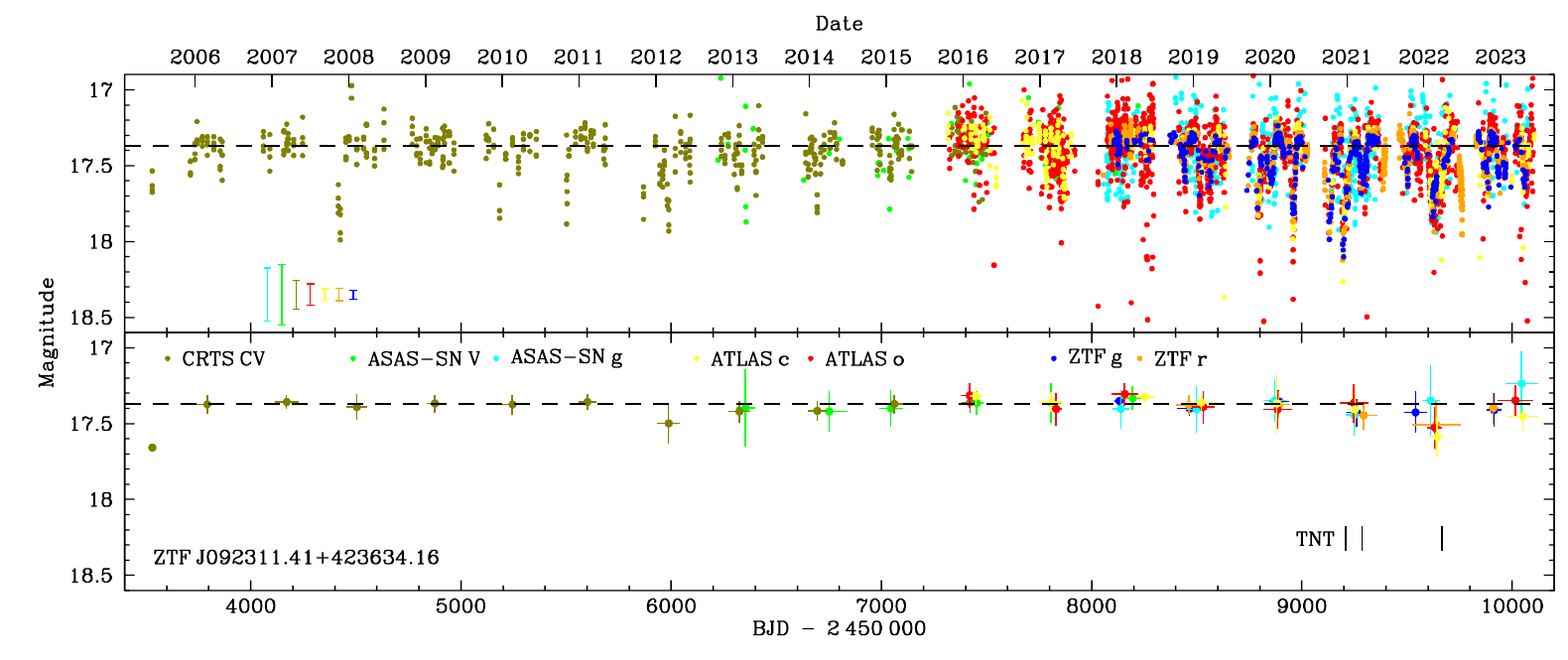}}
\caption{Long-term light curve of ZTF\,J0923+4236, all symbols have the same meaning as in Fig.\,\ref{fig:wdj1532_long}. Given the large-amplitude variability, the photometry of the different surveys were offset by their median values relative to the ZTF $g$-band data. It is apparent that the system has been undergoing deep ($\simeq1$\,mag) fading events over the past $\simeq17$\,yr which typically last a few months, see Fig.\,\ref{fig:0923_short} for a zoom-in on several of these events. Even though not strictly periodic, these events are sufficiently regular not to affect the seasonal median brightness of the system.  In contrast to ZTF\,J0328$-$1219 (Fig.\,\ref{fig:0328_long}), the seasonal magnitude of ZTF\,J0923+4236 has remained, within uncertainties, at a steady level, with the exception of the 2022/2023, where more scatter between the different surveys is seen.}
\label{fig:0923_long}
\end{figure*}

Noticeable is a large discrepancy between the seasonal averages of the CRTS and ASAS-SN $V$-band data in the three years where they overlap, 2012$-$2014. Given the sparse sampling of CRTS and ASAS-SN in 2012 and 2014 (in particular the first season of ASAS-SN $V$-band data, which only had four measurements), it is possible that the two surveys caught ZTF\,J0328$-$1219 in different states and/or different phases of the transiting debris. We note that the offsets determined from WD\,J1532+1356 may not be correct if the colour of ZTF\,J0328$-$1219 changes substantially with time, e.g. due to the production and subsequent dissipation of large amounts of small dust particles.

Our TNT light curves (Fig.\,\ref{fig:0328_tnt}) resemble those obtained by \citet{vanderboschetal21-1} (their Fig.\,2). Similar to the TNT light curves of WD\,1145+017 (e.g. Fig.\,1 in \citealt{gaensickeetal16-1}), ZTF\,J0328$-$1219 displays a mix of narrow, well-defined transit features and broader, complex absorption structures, with no constant ``out-of-transit'' stretch. As such, it is conceivable that the flux from the white dwarf is attenuated \textit{at all orbital phases}, and hence the star is dimmed overall with respect to its intrinsic brightness~$-$~which would be consistent with the fact that the latest long-term photometric data shows the brightness level of ZTF\,J0328$-$1219 still $\simeq0.1$\,mag fainter than during the earliest available survey data. This hypothesis would suggest that the transit activity, in terms of depth and/or number, was much higher during the 2011$-$2015 fading event. We analysed the CRTS and ASAS-SN $V$-band data obtained during those years using an Analysis-of-Variance (AoV) method \citep{schwarzenberg-czerny96-1}, but did not detect and periodic signal. Analysing the data from 2018 on-wards, we do detect the orbital period of 9.93\,h reported by \citet{vanderboschetal21-1}. The absence of a periodic signal during the dimming event may be the result of the sparse sampling of the CRTS and ASAS-SN $V$-band data, and/or relatively rapid evolution of the orbital period(s) of the debris causing the dimming.


\subsection{\protect{ZTF\,J092311.41+423634.16}}
ZTF\,J0923+4236 was identified by \citet{guidryetal20-1} as a white dwarf exhibiting irregular transits with complex structures in its ZTF light curve. A limited amount of fast photometry obtained by \citet{guidryetal20-1} revealed smooth, low-amplitude ($\simeq0.1$\,mag) variability with a time scale of $\simeq1$\,h. The authors classified the star from low-resolution spectroscopy as having a hydrogen-dominated (DA) atmosphere. 

\begin{figure}
\centering{\includegraphics[width=0.5\textwidth]{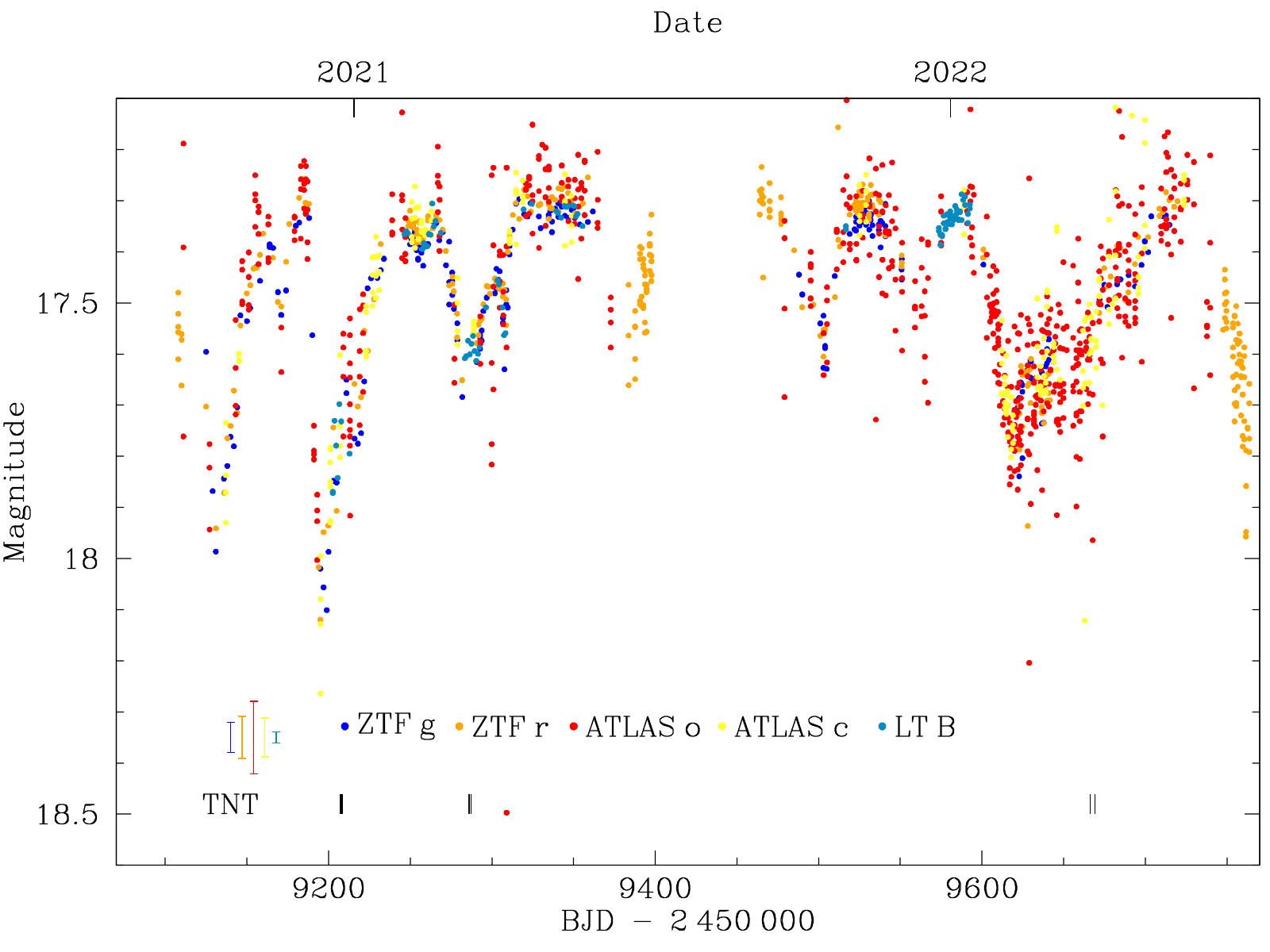}}
\caption{Zoom-in of the long-term light curve of ZTF\,0923+4236, all symbols have the same meaning as in Fig.\,\ref{fig:wdj1532_long}, with the addition of observations obtained with the LT.}
\label{fig:0923_short}
\end{figure}

The extended long-term light curve of ZTF\,J0923+4236 (Fig.\,\ref{fig:0923_long}, and Fig.\,\ref{fig:0923_short} for a zoom-in) demonstrates that the system has been undergoing fading events on time-scales of months, with an amplitude of up to $\simeq1$\,mag over the past 17 years. The seasonal average magnitudes remain broadly constant throughout the extended photometric coverage. We subjected the combined long-term light curve to an AoV analysis, but the only long-period signal detected is the lunar cycle.

\begin{figure}
\centering{\includegraphics[width=0.5\textwidth]{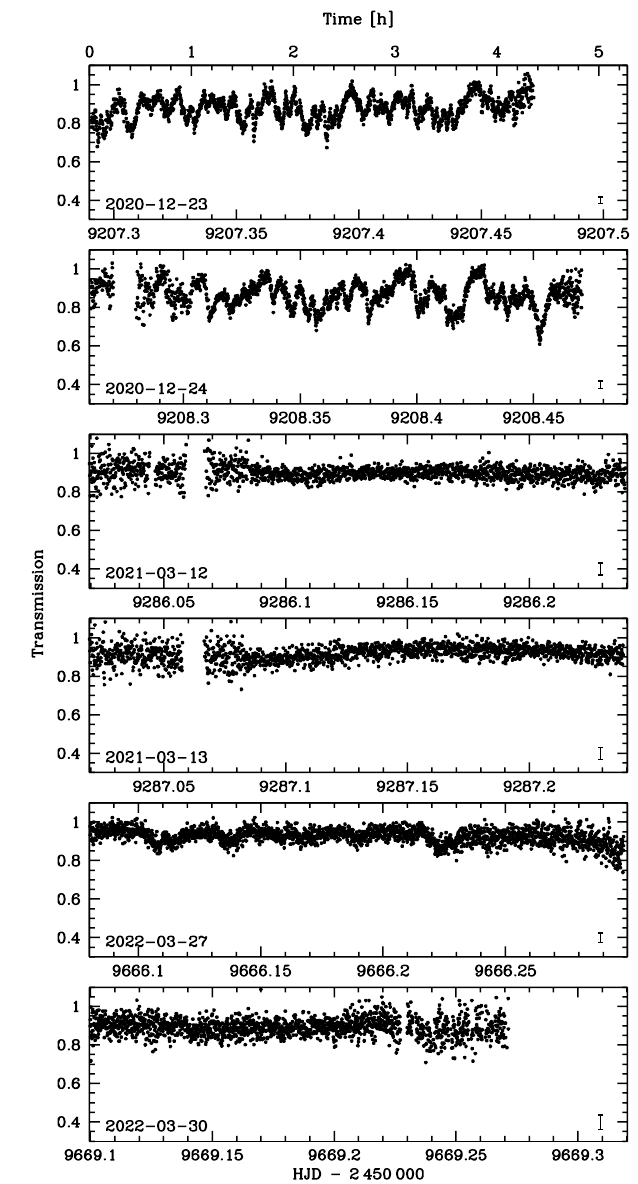}}
\caption{TNT light curves of ZTF\,0923+4236. The transit activity appears to be correlated with the overall brightness, the light curves in the top two panels were obtained near the faintest state of the system (Fig.\,\ref{fig:0923_short}), in contrast, the middle two light curves were obtained during an intermediate brightness state. The slight modulation in the 2021 March 13 light curve likely arises from an imperfect extinction correction.  The median values of the photometric uncertainties are illustrated by the error bar in the bottom-right of each panel. }
\label{fig:0923_tnt}
\end{figure}

We obtained the first TNT data of ZTF\,J0923+4236 in December 2020, when the system was near its faintest state (see Fig.\,\ref{fig:0923_short}), and the light curves display rapid variability on time scales of a few minutes (Fig.\,\ref{fig:0923_tnt}). We analysed both nights separately, and whereas the AoV power spectra show multiple signals in the range of $\simeq10$\,min to $\simeq$1\,h, none of these signals appear in both nights, and we conclude that the observations contain no evidence for a periodic signal. Overall, these TNT light curves are reminiscent of the transiting WD\,1054$-$226 (\citealt{farihietal22-1}, e.g. their Fig.\,2 and 6), though with variability occurring on even shorter time scales in ZTF\,J0923+4236. It is important to keep in mind that transiting debris can only result in dimming, not in brightening. Hence, the true flux level of the unobscured white dwarf is likely to be higher than the brightest segments of these  two light curves, i.e. absorption by transiting debris blocks some amount of the white dwarf flux at all orbital phases. 

When we re-visited ZTF\,J0923+4236 in March 2021, the system was in a slightly brighter state, and the two TNT light curves are totally devoid of any transit activity. Further observations obtained in March 2022, again at an intermediate brightness level, show mild variability on time scales of $\simeq1$\,h, similar to the light curves shown by \citet{guidryetal20-1} (their Fig.\,10), which faded away in our last observation. 

Whereas the larger transit activity observed in December 2020 will result in an overall dimming of the average brightness of the white dwarf, the long-term light curve (Fig.\,\ref{fig:0923_short}) shows an overall fading of nearly one magnitude, larger than the maximum depth of $\simeq20$~per cent observed on short time scales within the TNT light curves. It thus appears that the transit activity correlates with an overall dimming of the system. The absence of a periodicity in the long-term variability is very puzzling, as a debris stream on an eccentric orbit would result in periodic dimming, as observed every $\simeq107$\,d in ZTF\,J0139+5245 \citep{vanderboschetal20-1}.

\begin{figure}
\centering{\includegraphics[width=0.5\textwidth]{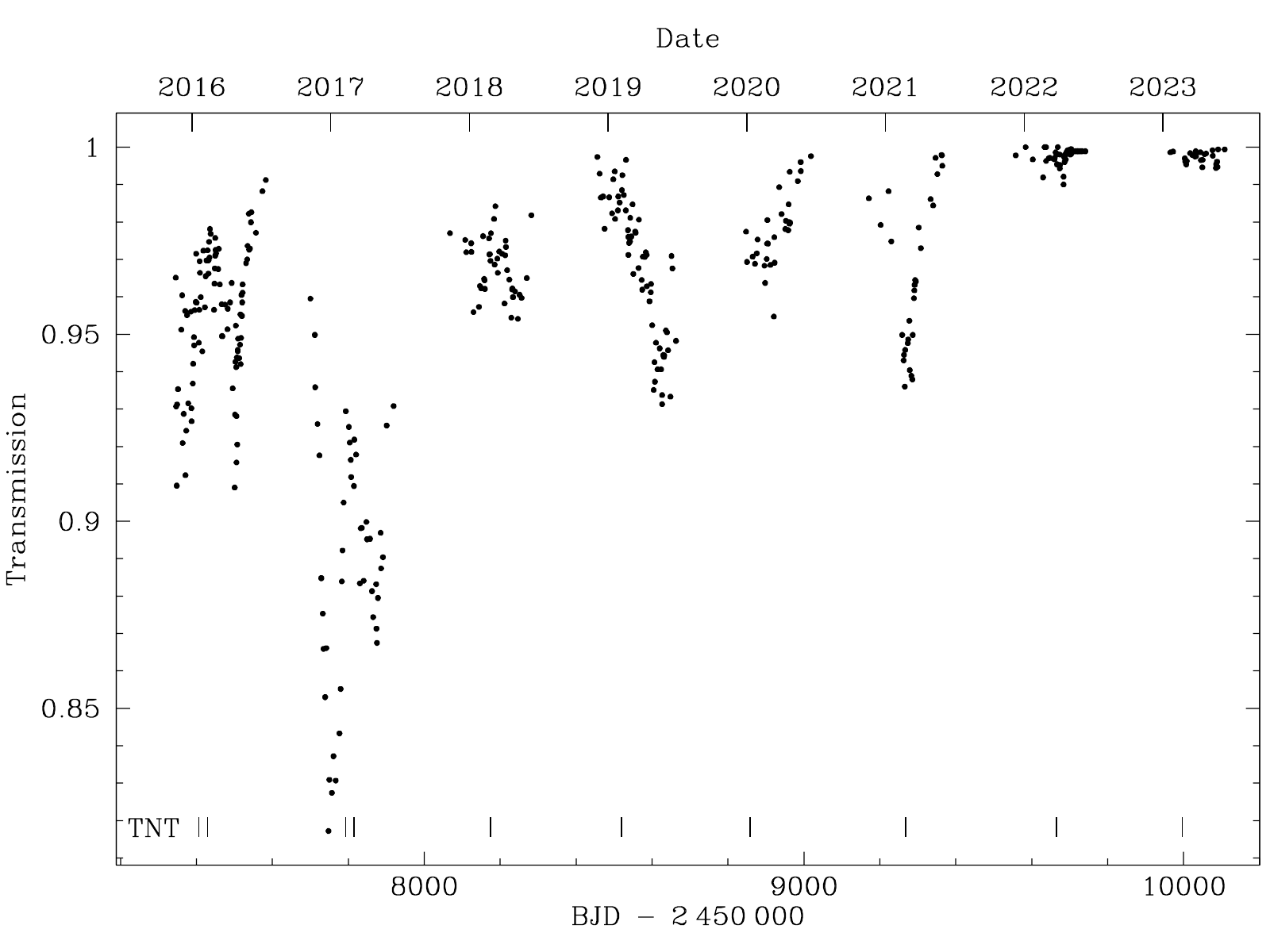}}
\caption{Long-term averages of the orbital transmission of WD\,1145+017, derived from intense photometric monitoring of the system using the facilities as described by \citet{garyetal17-1}.}
\label{fig:1145_bruce}
\end{figure}

\subsection{WD\,1145+017}
WD\,1145+017 is the prototype of white dwarfs exhibiting debris transits, discovered by \citet{vanderburgetal15-1} using \textit{K2} data. \citet{vanderburgetal15-1} determined an dominant orbital period of 4.49\,h, and additional signals at slightly shorter orbital periods. The discovery of this system triggered a flurry of photometric \citep{gaensickeetal16-1, rappaportetal16-1, crolletal17-1, garyetal17-1}, spectroscopic \citep{xuetal16-1, redfieldetal17-1, hallakounetal17-1, izquierdoetal18-1, karjalainenetal19-1, fortin-archambaultetal20-1}, X-ray and polarimetric \citep{farihietal18-1} and theoretical \citep{gurrietal17-1, verasetal17-2, duvvurietal20-1} studies, providing very detailed insight into the morphology of the debris transits, and into the physical processes occurring at WD\,1145+017. 

\begin{figure*}
\centering{\includegraphics[width=1.0\textwidth]{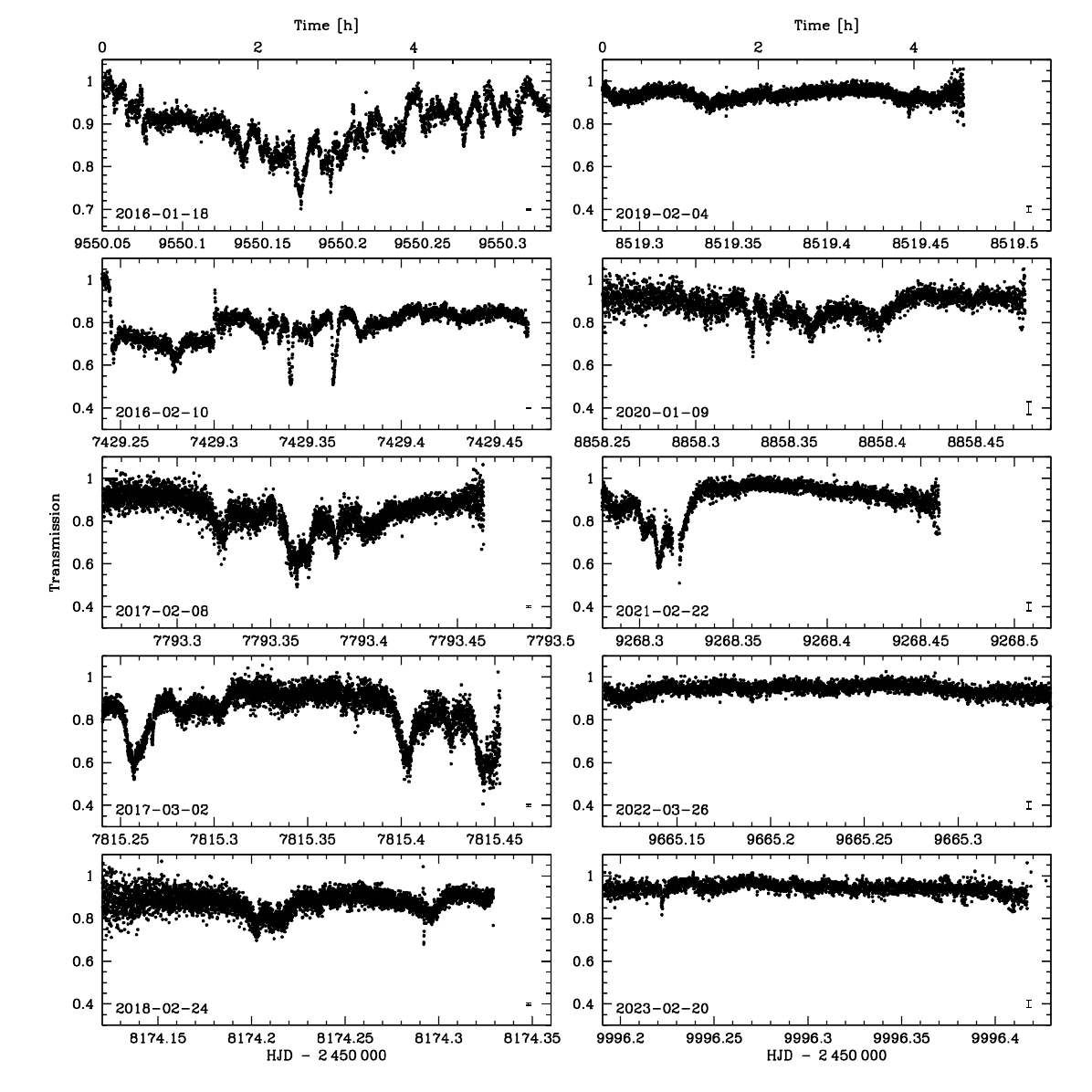}}
\caption{TNT light curves of WD\,1145+017 obtained from 2016 to 2023 illustrate the large variation in the transit activity of the system, see Fig.\,\ref{fig:1145_bruce} for detailed monitoring of the orbital averaged transmission throughout the same time span. Most of the light curves displayed here cover the dominant 4.49\,h orbital period of the debris. The median values of the photometric uncertainties are illustrated by the error bar in the bottom-right of each panel. }
\label{fig:1145_tnt}
\end{figure*}

\citet{garyetal17-1} discussed the changes in transit activity of WD\,1145+017, based on intense photometric monitoring of the system from December 2015 throughout July 2016. The authors defined a photometric equivalent width as the orbital average of the transit absorption, and showed (their Fig.\,9 and 10) that this measure increased from a fraction of a per cent during the \textit{K2} observations to reaching nearly ten per cent in 2016, with large fluctuations on time scales of months. 

Here, we present an update of the analysis of \citet{garyetal17-1}, including additional observations from mid-2016 to July 2023. We use here a slightly different definition, showing the data as transmission, which is essentially one minus the equivalent width defined by \citet{garyetal17-1}. The new data (Fig.\,\ref{fig:1145_bruce}) shows that the overall transit activity reached a maximum around 2017, with the average transmission dropping almost to 80~per cent. In the following years, the transit activity decreased, with the average transmission varying from 2019 to 2021 between close to zero and five per cent. After that, the average transmission remained at very high levels of about one per cent. 

\begin{figure*}
\centering{\includegraphics[width=1.0\textwidth]{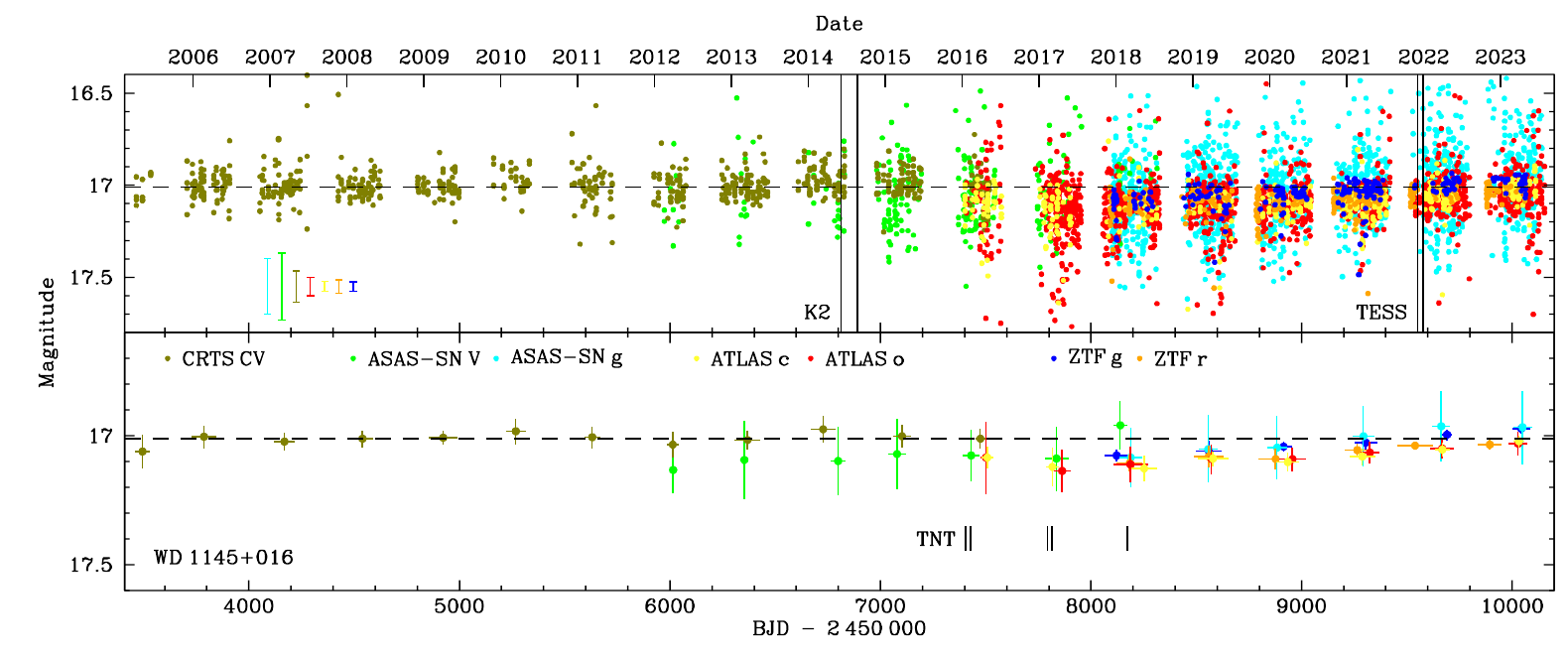}}
\caption{Long-term light curve of WD\,1145+017, all symbols have the same meaning as in Fig.\,\ref{fig:wdj1532_long}. We have used WD\,J093204.05+285044.61 as reference star to compute offsets between the different surveys. The \textit{K2} observations that led to the dicovery of the debris transits in WD\,1145+017 were obtained in between the two vertical lines shown in the top panel. That period has only been covered by relatively sparse ASAS-SN $V$-band observations, however, a steady brightening of the system is seen in the seasonal data of the ASAS-SN $g$-band as well as in both bands of ZTF and ATLAS. This is consistent with the increase of the average transmission measured from detailed time-series photometry of the system, see Fig.\,\ref{fig:1145_bruce}. \textit{TESS} observations were obtained during a quiescent period (see the last two panels of Fig.\,\ref{fig:1145_tnt}).}
\label{fig:1145_long}
\end{figure*}

We have obtained fast TNT photometry of WD\,1145+017 from November 2015 throughout April 2023, which illustrates the change in the transit activity of the system, both in terms of morphology of the individual transits and the overall light curve (Fig.\,\ref{fig:1145_tnt}). During very active periods (2016 to 2017), there is hardly any segment of the light curve that is not affected by transits, with the implication that the orbital average transmission derived from these data should be considered a lower limit. In 2018 and 2019, the deep and sharp transits changed into broader and shallower absorption structures, consistent with azimuthal spreading of the obscuring material. Our latest data obtained in 2022 and 2023 exhibits nearly no transit activity, apart from one single short transit detected on 2023 February 20. Such sharp transits are likely related to dust freshly released from a fragment either undergoing a collision with another solid body, or suffering structural stress due to tidal effects.

The long-term light curve of WD\,1145+017 (Fig.\,\ref{fig:1145_long}) shows the system $\simeq0.1$\,mag fainter in the seasonal averages in the ASAS-SN $V$-band, which is within the limits of our offset calibration for that specific band. However, the ASAS-SN $V$-band data agrees well with the average ZTF $r$-band and ATLAS $c$-band data obtained in 2016 and 2017, suggesting that the drop in the ASAS-SN $V$-band brightness is real. This is somewhat puzzling, as the \textit{K2} data presented by \citet{vanderburgetal15-1}, obtained in the middle of the ASAS-SN $V$-band survey, caught the system in a state of low transit activity, and the increase in activity was only recorded shortly after the \textit{K2} observations (e.g. \citealt{gaensickeetal16-1, rappaportetal16-1}). However, the calibration of the offsets of the survey pass-bands is colour dependent, so it is possible that WD\,1145+017 might have changed both in brightness and/or colour around 2012 or 2013. As mentioned earlier, a small colour-dependence of the transit depths was found by \citet{hallakounetal17-1}. Whereas the observed amplitude of that colour dependence is insufficient to explain the long-term trend discussed above, our current insight into the evolution of the debris transits is insufficient to rule out larger colour variations at certain epochs.

The period 2016 to 2023 is covered by multiple photometric surveys, and the seasonal averages agree well. All the survey data indicates that the system has been brightening by $\simeq0.1$\,mag from 2017 to 2023, which is consistent with the decrease of the transit activity, and the increase of the average orbital transmission discussed above.  

\begin{figure}
\centering{\includegraphics[width=0.5\textwidth]{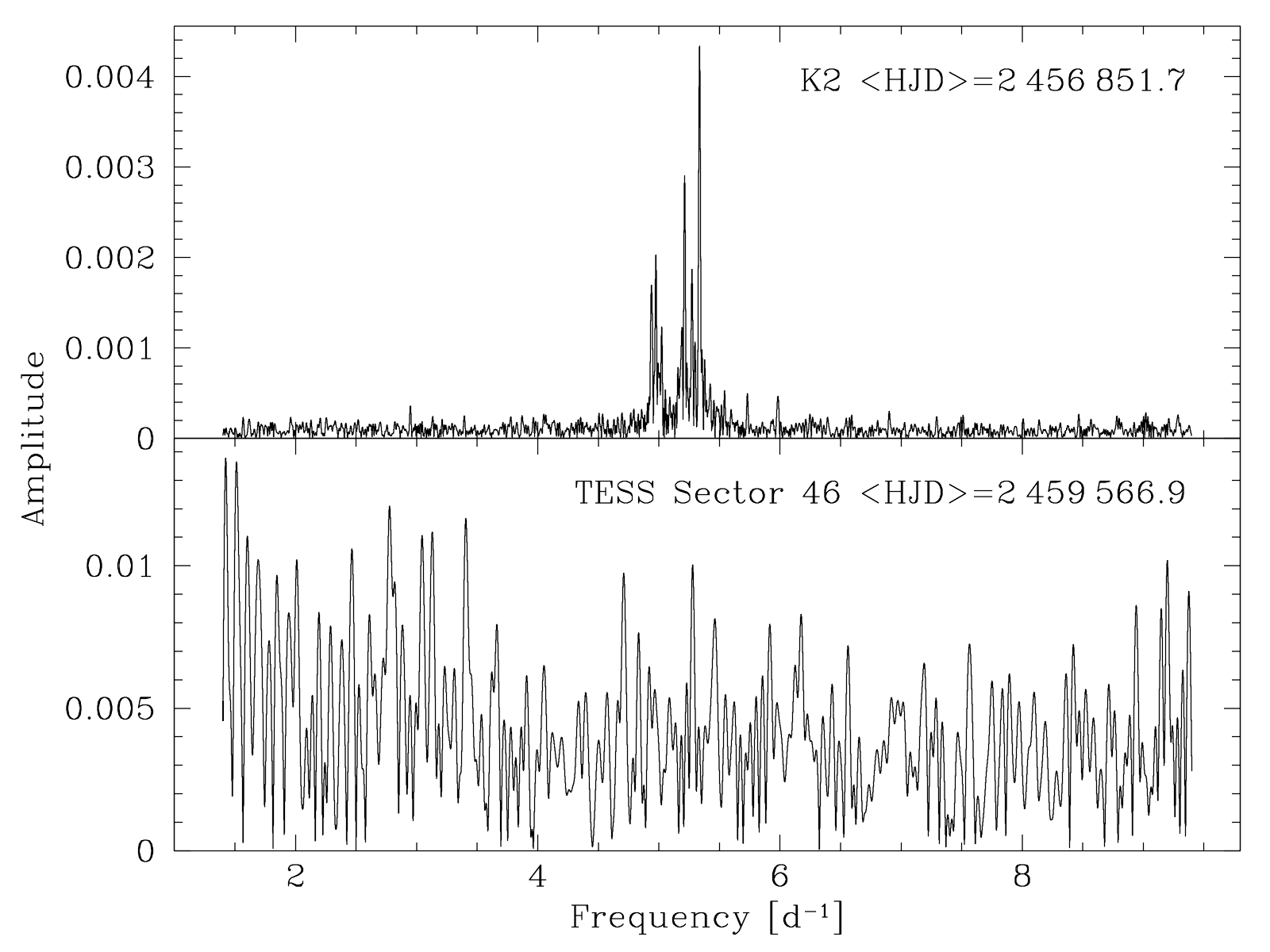}}
\caption{Amplitude spectra computed from the \textit{K2} and \textit{TESS} Sector~46 observations, the mean HJD epochs of the two data sets are labelled. To compare the two data sets, we normalised the light curves to a median flux of one prior to computing the amplitude spectra. The \textit{K2} data clearly reveals the periodic transit signals identified by \citet{vanderburgetal15-1}, which are not present in the \textit{TESS} observations.}
\label{fig:1145_power_k2_tess}
\end{figure}

\textit{TESS} observations of WD\,1145+017 were obtained in Sector~46 between 2021 December 4 to December 31. We retrieved the \textit{TESS} short-cadence (20\,s) pre-search data conditioning simple aperture (PDCSAP) fluxes, as well as the \textit{K2} PDCSAP fluxes, and computed discrete Fourier transforms for both data sets.  A detailed comparison of the transit detection efficiency of the two data sets is difficult, as they were taken with different aperture sizes (\textit{K2}: 0.95\,m, \textit{TESS}: 0.11\,m), cadences (\textit{K2}: 29.4\,min, \textit{TESS}: 20\,s), and total baselines (\textit{K2}: 80\,d, \textit{TESS}: 25.7\,d). The crowding in \textit{TESS} is not very severe, WD1145+017 has a CROWDSAP value of 0.93 in Sector 46, suggesting that only about seven~per cent of the flux in the aperture is not coming from the white dwarf. This has been incorporated in the relative fluxes in the PDCSAP \textit{TESS} light curve. As a simple approach, we computed the amplitude spectra of the \textit{K2} and \textit{TESS} data after normalising both light curves to a median flux of one. Whereas the amplitude spectrum of the \textit{K2} data clearly shows the multiple transit signals identified by \citet{vanderburgetal15-1}, they are not detected in the \textit{TESS} data (Fig.\,\ref{fig:1145_power_k2_tess}). This non-detection of a periodic signal in the \textit{TESS} data is consistent with very weak or absent transits in the TNT light curves obtained in 2022 and 2023 (Fig.\,\ref{fig:1145_tnt}).

\section{Discussion}
We detect long-term variations in the brightness of three white dwarfs exhibiting transiting debris: ZTF\,J0328$-$1219, ZTF\,J0923+4236 and WD\,1145+017. Taking the average of the CRTS and ASAS-SN $V$-band data, ZTF\,J0328$-$1219 faded by $\simeq0.3$\,mag around 2012 to 2014, and has not yet fully recovered to the brightness level recorded by the earliest CRTS data. This might suggest that the system underwent a large collisional event around 2011, resulting in the production of large amounts of dust occulting the white dwarf, which has since then gradually dispersed, though leaving sufficient material to account for the ongoing transit activity, which imply continued dust production. 

Similarly, WD\,1145+017 exhibits an overall drop in its brightness which concurs with a large increase in transit activity, followed by a subsequent gradual re-brightening. In the case of WD\,1145+017, the long-term variations seen in the photometric survey data is accompanied by very detailed time-series photometry, corroborating that the overall trends seen in the brightness of WD\,1145+017 are linked to varying amounts of transit activity. 

The most puzzling case is that of ZTF\,J0923+4236, which exhibits deep aperiodic fading events over the past 17 years, and large changes in the level of short-term variability which appears to anti-correlate with the overall brightness of the system. 

These long-term changes may be the result of the ongoing disruption of a planetesimal, or the collision between multiple fragments, both leading to a temporarily increased dust production. Such an increased amount of dust production may result in a brightening at infrared wavelengths, as has been observed in WD\,0145+234 \citep{wangetal19-1}.

Searching for photometric transit signals at white dwarfs is a very active research field \citep[e.g.][]{faedietal11-1, fultonetal14-1, sandhausetal16-1, belardietal16-1, dameetal19-1}, resulting, however, so far in very few detections: currently, only eight white dwarfs are known, or suspected to exhibit transits from planetary debris \citep{vanderburgetal15-1, vanderboschetal20-1, guidryetal20-1, farihietal22-1}. However, it is now clear that these searches will be affected by variations in transit activity~--~WD\,1145+017 in its current state would be very unlikely identified from ground-based time-series photometry (Fig.\,\ref{fig:1145_bruce} and Fig.\,\ref{fig:1145_tnt}, bottom-right two panels). 

This makes an assessment of the frequency of white dwarfs hosting disintegrating planetesimals difficult.  \citet{vansluijs+vaneylen17-1} analysed the light curves of 1148 white dwarfs observed by \textit{Kepler/K2} for transit signals, and did not discovery any system containing a disintegrating planetesimal beyond WD\,1145+017 \citep{vanderburgetal15-1}. The \textit{Kepler/K2} white dwarf sample is a random mix drawn from the literature, broadly speaking, magnitude limited, biased towards younger, brighter systems, and hence spanning a wide range of distances (WD\,1145+017 is at $d\simeq145$\,pc, \citealt{gaiaetal20-1})~--~and it implies an detection rate of debris transits of $\simeq1/1000$. Among the eight confirmed and candidate systems with debris transits, one is within 40\,pc (WD\,1054$-$226, $d=36$\,pc, \citealt{farihietal22-1}), and considering that there are $\simeq1000$ white dwarfs within 40\,pc \citep{mccleeryetal20-1, tremblayetal20-1, obrienetal23-1} again suggests a rate of $\simeq1/1000$~--~however, the 40\,pc sample has not yet been homogeneously surveyed for transit signals. The closest white dwarf exhibiting debris transits reported by \citet{guidryetal20-1} is ZTF\,J0328$-$1219 ($d=43$\,pc), just slightly outside the 40\,pc sample. \citet{guidryetal20-1} applied a temperature cut of $7000-16\,000$\,K in their analysis, as well as astrometric and photometric quality cuts, which reduced the number of white dwarfs in their analysis with $d\lesssim40$\,pc to $\simeq450$~--~of which only about half have ZTF light curves, suggesting a detection rate of $\simeq1/200$. 

The details of the observational biases in the \textit{Kepler/K2}, 40\,pc and \citet{guidryetal20-1} samples are difficult to quantify and compare, even more so as the case of WD1145+017 demonstrates that the strength of photometric transit signals can strongly vary on time scales of years. Nevertheless, these three independent estimates, crude as they are, suggest that white dwarfs with detectable debris transits are not exceedingly rare. Factoring in that the detection implies a highly aligned orbital inclination, the true number of white dwarfs hosting planetesimals that are currently disintegrating is factors of several tens higher. It is hence conceivable that the fraction of white dwarfs with close-in debris undergoing disintegration is within factors of ten to a few of the fraction of white dwarfs exhibiting photospheric metals, which is $\simeq25$~per cent of all white dwarfs, and the unmistakable signature of ongoing or recent accretion \citep{zuckermanetal03-1, zuckermanetal10-1, koesteretal14-1}. With only eight confirmed and candidate systems, it is still too early to gauge whether the intrinsic fraction of these systems among all white dwarfs might correlate with their effective temperatures, and hence cooling ages and/or their masses.  

Extending the same reasoning to the transiting planet candidate at WD\,1856+534 with $d\simeq25$\,pc \citep{vanderburgetal20-1} results in a detection rate of $\sim1/250$, which, again because of the need for a closely aligned viewing geometry, is highly under-estimating the true number of white dwarfs with planets on close-in orbits. Other relatively nearby white dwarfs with subtle photometric or spectroscopic potential signatures of non-transiting planets (GD\,394 at $d\simeq50$\,pc \citealt{dupuisetal00-1, wilsonetal19-1, wilsonetal20-1}; GD\,356 at $d\simeq20$\,pc, \citealt{lietal98-1, gaensickeetal20-2}; WD\,J122619.77+183634.46 at $\simeq36$\,pc, \citealt{manseretal23-1}, and WD\,J041246.84+754942.26 at $d\simeq35$\,pc and WD\,J165335.21$-$100116.33 at $d\simeq33$\,pc, \citealt{elmsetal23-1}) corroborate the hypothesis that these systems might be relatively common.

\section{Conclusions}
We detect long-term variability in three white dwarfs that are known to exhibit transits from planetary debris, and at least in two cases, the long-term changes in brightness correlate with the level of activity detected in high-speed photometry on a timescale of several hours. Our study highlights the potential of a joint analysis of the nearly two decades of large-area survey photometry as a novel approach in identifying white dwarfs that exhibit irregular variability. The scientific potential of future time-domain surveys would benefit from (quasi)simultaneous multi-band photometry to probe for changes in brightness and colour. Based on crude, but independent estimates, we suggest that debris transits may be detectable among one in a few hundred white dwarfs, and that it is likely that a significant fraction of all white dwarfs exhibiting photospheric metals currently host disintegrating planetesimals. 

\section*{Acknowledgements}
We thank the referee for a detailed and constructive report. 
We are grateful to Bruce Gary for his continued observations, analysis, and interpretation of WD\,1145+017, and for contributing his observations to this project.
This research has received funding support from the National Science, Research and Innovation Fund (NSRF) via the Program Management Unit for Human Resources \& Institutional Development, Research and Innovation (Grant No. B05F640046).
This project has received funding from the European Research Council (ERC) under the European Union’s Horizon 2020 research and innovation programme (Grant agreement No. 101020057).
This research was supported in part by the National Science Foundation under Grant No. PHY-1748958.
This research was supported by Thailand Science Research and Innovation (TSRI) (Grant no. R2565B083).
This work has made use of data obtained at the Thai National Observatory on Doi Inthanon, operated by NARIT.
The CSS survey is funded by the National Aeronautics and Space Administration under Grant No. NNG05GF22G issued through the Science Mission Directorate Near-Earth Objects Observations Program.  The CRTS survey is supported by the U.S.~National Science Foundation under grants AST-0909182.
Based on observations obtained with the Samuel Oschin 48-inch Telescope at the Palomar Observatory as part of the Zwicky Transient Facility project. ZTF is supported by the National Science Foundation under Grant No. AST-1440341 and a collaboration including Caltech, IPAC, the Weizmann Institute for Science, the Oskar Klein Center at Stockholm University, the University of Maryland, the University of Washington, Deutsches Elektronen-Synchrotron and Humboldt University, Los Alamos National Laboratories, the TANGO Consortium of Taiwan, the University of Wisconsin at Milwaukee, and Lawrence Berkeley National Laboratories. Operations are conducted by COO, IPAC, and UW. 
The Pan-STARRS1 Surveys (PS1) and the PS1 public science archive have been made possible through contributions by the Institute for Astronomy, the University of Hawaii, the Pan-STARRS Project Office, the Max-Planck Society and its participating institutes, the Max Planck Institute for Astronomy, Heidelberg and the Max Planck Institute for Extraterrestrial Physics, Garching, The Johns Hopkins University, Durham University, the University of Edinburgh, the Queen's University Belfast, the Harvard-Smithsonian Center for Astrophysics, the Las Cumbres Observatory Global Telescope Network Incorporated, the National Central University of Taiwan, the Space Telescope Science Institute, the National Aeronautics and Space Administration under Grant No. NNX08AR22G issued through the Planetary Science Division of the NASA Science Mission Directorate, the National Science Foundation Grant No. AST-1238877, the University of Maryland, Eotvos Lorand University (ELTE), the Los Alamos National Laboratory, and the Gordon and Betty Moore Foundation.
This paper includes data collected by the Kepler mission and obtained from the MAST data archive at the Space Telescope Science Institute (STScI). Funding for the Kepler mission is provided by the NASA Science Mission Directorate. STScI is operated by the Association of Universities for Research in Astronomy, Inc., under NASA contract NAS 5–26555.
This paper includes data collected by the TESS mission, which are publicly available from the Mikulski Archive for Space Telescopes (MAST). Funding for the TESS mission is provided by NASA's Science Mission directorate.
\section*{Data Availability}
The ASAS-SN, ATLAS and ZTF light curves can be retrieved from the respective archives, and we will make the CRTS and TNT light curves available upon reasonable requests communicated to the lead author.



\bibliographystyle{mnras}
\bibliography{aabib} 





\bsp	
\label{lastpage}
\end{document}